\documentclass[12pt]{article}
\usepackage{geometry}                		
\usepackage[parfill]{parskip}    		
\usepackage{graphicx}				
\usepackage{color}								
\usepackage{amssymb}
\usepackage{amsmath}
\usepackage{amssymb}
\usepackage{graphicx}
\usepackage{placeins}
\usepackage[subfigure]{tocloft}
\usepackage{subfigure}
\usepackage{float}
\usepackage{fullpage}
\usepackage{braket}
\usepackage{url}
\usepackage{mcite}
\usepackage{upgreek}
\usepackage{outlines}
\usepackage{braket}
\usepackage{appendix}
\usepackage{xspace}
\usepackage{multirow}
\usepackage{authblk}
\usepackage{mathtools}
\usepackage{tikz}
\usepackage{hyperref}
\usetikzlibrary{matrix,arrows,decorations.pathmorphing}
\usetikzlibrary{patterns}
\usetikzlibrary{trees}
\usetikzlibrary{decorations.pathmorphing}
\usetikzlibrary{decorations.markings}
\usetikzlibrary{positioning,arrows}
\usetikzlibrary{decorations.pathreplacing}
\usetikzlibrary{decorations.pathmorphing}
\usetikzlibrary{decorations.markings}
\tikzset{
particle/.style={thin,draw=black, postaction={decorate},
decoration={markings,mark=at position .5 with {\arrow[black, line width=0.5mm]{stealth}}}},
gluon/.style={decorate, draw=black, decoration={snake=coil}}
}
\tikzstyle{pomeron} = [thin,draw=black]
\tikzstyle{anti} = [thin,draw=black, postaction={decorate},
decoration={markings,mark=at position .5 with {\arrow[black, line width=0.5mm]{stealth reversed}}}]
\tikzstyle{antivirasoro} = [thin,draw=black, postaction={decorate},
decoration={markings,mark=at position .333 with {\arrow[black, line width=0.5mm]{stealth reversed}}}]
\tikzstyle{antivirasoro2} = [thin,draw=black, postaction={decorate},
decoration={markings,mark=at position .666 with {\arrow[black, line width=0.5mm]{stealth reversed}}}]
\tikzstyle{virasoroparticle} = [thin,draw=black, postaction={decorate},
decoration={markings,mark=at position .333 with {\arrow[black, line width=0.5mm]{stealth}}}]
\tikzstyle{virasoroparticle2} = [thin,draw=black, postaction={decorate},
decoration={markings,mark=at position .666 with {\arrow[black, line width=0.5mm]{stealth}}}]
\tikzstyle{arrow} = [thin,draw=black, postaction={decorate},
decoration={markings,mark=at position 1 with {\arrow[black, line width=0.5mm]{stealth}}}]
\tikzstyle{particlec1} = [thin,draw=black, postaction={decorate},
decoration={markings,mark=at position .25 with {\arrow[black, line width=0.5mm]{stealth}}}]
\tikzstyle{particlec2} = [thin,draw=black, postaction={decorate},
decoration={markings,mark=at position .75 with {\arrow[black, line width=0.5mm]{stealth}}}]
\tikzstyle{particlec3} = [thin,draw=black, postaction={decorate},
decoration={markings,mark=at position .95 with {\arrow[black, line width=0.5mm]{stealth}},mark=at position .05 with {\arrow[black, line width=0.5mm]{stealth}}}]

\newcommand\ignore[1]{}

\newcommand\be{\begin{equation}}
\newcommand\eeq{\end{equation}}
\newcommand\bea{\begin{eqnarray}}
\newcommand\eea{\end{eqnarray}}

\newcommand{\bdm}{\begin{displaymath}}
\newcommand{\edm}{\end{displaymath}}

\newcommand{\dslash}[2]{{{#1}\hspace{-5pt}{/}}_{#2}}

\renewcommand\t[1]{\tilde{#1}}

\renewcommand{\epsilon}{\varepsilon}
\renewcommand{\phi}{\varphi}
\DeclareMathOperator{\Disc}{Disc}

\numberwithin{equation}{section}
\numberwithin{figure}{section}

\def\Dslash{D \!\!\!\! \slash \;}

\def\zm{{z_{max}}}
\def\k{\kappa}

\def\pt{p_{\mathrm{T}}}
\renewcommand\t[1]{\tilde{#1}}

\def\brown{Department of Physics\\
Brown University, 182 Hope St. Providence, RI 02902 USA}
\def\kansas{Department of Physics and Astronomy\\
University of Kansas, Lawrence, KS 66045 USA}
\def\stanford{Department of Physics\\
Stanford University, Stanford, California 94305 USA}
\def\supportk{\footnote{Work supported by Department of Energy under contact DE-Sc0010010-Task-A and the Foundation Distinguished Professor starting grant of Prof. Christophe Royon}}
\def\supportb{\footnote{{\bf Speaker.} Work supported by Department of Energy under contact DE-Sc0010010-Task-A}}
\def\supports{\footnote{Work supported by Stanford Phys. Dept., Stanford EDGE grant, and NSF Fellowship number DGE-1656518}}

\def\Title#1{\begin{center} {\Large #1 } \end{center}}
\def\Author#1{\begin{center}{ \sc #1} \end{center}}
\def\Address#1{\begin{center}{ \it #1} \end{center}}

\newenvironment{Abstract}{\begin{quotation}  }{\end{quotation}}
\newenvironment{Presented}{\begin{quotation} \begin{center} 
             PRESENTED AT\end{center}\bigskip 
      \begin{center}\begin{large}}{\end{large}\end{center} \end{quotation}}

\def\beq{\begin{equation}}
\def\eeq#1{\label{#1}\end{equation}}
\def\eeqn{\end{equation}}

\def\beqa{\begin{eqnarray}}
\def\eeqa#1{\label{#1}\end{eqnarray}}
\def\eeqan{\end{eqnarray}}

\let\bar=\overbar

\def\Dslash{\not{\hbox{\kern-4pt $D$}}}
\def\dslash{\not{\hbox{\kern-2pt $\del$}}}

\def\msb{{\bar{\ssstyle M \kern -1pt S}}}

\begin{document}
\begin{titlepage}

\vfill
\Title{Holographic Inclusive Central Particle Production at the LHC}
\vfill
\Author{ $^{\#}$ Richard Nally\supports}
\Author{ $^*$Timothy Raben\supportk}
\Author{ {\bf ${\dagger}$}{\bf Chung-I Tan}\supportb}
\Address{$^{\#}$\stanford}
\Address{$^*$\kansas}
\Address{{\bf ${\dagger}$}\brown}
\vfill
\begin{Abstract}
We describe how the holographic, BPST Pomeron can be used to describe inclusive central production at the LHC.
\end{Abstract}
\vfill
\begin{Presented}
Presented at EDS Blois 2017, Prague, Czech Republic, June 26-30,2017
\end{Presented}
\vfill
\end{titlepage}
\def\thefootnote{\fnsymbol{footnote}}
\setcounter{footnote}{0}


\section{Introduction}\label{sec:Intro}
The AdS/CFT has provided new avenues for investigating hadron behavior in the non-perturbative regime.  It has become a novel tool for understanding how conformal theories can describe scattering experiments.  Here we present a summary of a comprehensive treatment~\cite{Nally:2017nsp} of a conformal theory describing central inclusive production at the LHC. Conformal scattering can be described as the flow of infrared safe observables\ignore{~\cite{Sterman:1977wj,Brown:1981jv},} leading to vacuum expectations
\begin{equation} \label{eq:Hofman1}
\sigma_{w}(p)=\int d^4x e^{-ipx}  \langle 0| {\cal O}^\dagger(x) {\cal D}[w] {\cal O} (0) |0\rangle,
\end{equation}
where ${\cal O}$ is the source for an initial state. The inclusive event shape distribution ${\cal D}[w]$, a product of a set of local operators, measures flows of conserved  quantities.

In momentum space, generalized optical theorems identify inclusive cross sections as \emph{discontinuities} of appropriate forward  amplitudes. For example, the differential cross section $ {d \sigma_{ab\rightarrow c+X}}/{(d^3{\bf p}_c/E_c)}$  of the process   $ab\rightarrow c +X$  can  be identified as the discontinuity in $M^2$ of the amplitude for the six-point process $abc'\rightarrow a'b'c$.  This process can be described holographically where a high energy scattering process depends crucially on the conformal data. Symbolically, we have 
\begin{equation}
 \frac{d \sigma_{ab\rightarrow c+X}}{d^3{\bf p}_c/E_c} \propto \frac{1}{2i s}  {\rm Disc}_{M^2} T_{abc'\rightarrow a'b'c} \xrightarrow[\text{AdS/CFT}]{\text{Regge Limit}} {\rm Disc}_{M^2}\langle {\cal V}_P V_{c\bar c} {\cal V}_P\rangle
 \sim p_\perp^{-\delta} \, ,  \label{eq:Optical6}
\end{equation}
where $\delta$ is related to a conformal scaling dimension. Here we review the key results needed to arrive at Eq.(\ref{eq:Optical6}) and apply it to scattering at the LHC.  Further details and examples can be found in ~\cite{Nally:2017nsp}.

\section{Inclusive Cross-Sections and Discontinuities}\label{sec:disc}
In general, n-point momentum space Wightman correlation functions are related to the forward discontinuity of the associated n-point \emph{time-ordered} Green's function\footnote{The optical theorem being the simplest example}. For example, the invariant single particle inclusive differential cross section, $d\sigma_{ab\rightarrow x+X}/d^3{\bf p}_c/E_c$, can be expressed as~\cite{Mueller}
\begin{equation}
 d \sigma \propto \sum_X (2\pi)^4 \delta^{(4)}(p_a+p_b-p_c-p_X) \Big| \langle p_c,X \Big| p_a,p_b\rangle \Big|^2 \propto    \langle p_{a},p_{b}|   \widetilde {\cal O}_c(p_c)  |p_a,p_b\rangle \,.    \label{eq:Optical6a}
\end{equation}
Here  $\widetilde {\cal O}_c =    \int d^4 x e^{-ip_c\cdot x}     \phi_c(x)  \phi_c(0)  $ is the Fourier transform of   product of two local operators. Since $ \phi_c(x)  \phi_c(0)$ is not time-ordered, one is dealing with a Wightman function.  
\ignore{
Motivated by the work of Hofman and Maldacena \cite{Hofman:2008ar}, a similar analysis holds for processes with a single local source. For example, consider virtual photon decay. Here we can recast Eq (\ref{eq:Hofman1}) in the form of  a normalized distribution in a momentum representation as 
\begin{equation} \label{eq:Hofman2}
\langle \widetilde  O_w\rangle =\frac{\sigma_{w}(p)}{\sigma_{\cal O}(p)} =\frac{\int d^4x e^{ipx}  \langle 0| {\cal O}^\dagger(x) \widetilde  O_w {\cal O} (0) |0\rangle} {\int d^4x e^{ipx}  \langle 0| {\cal O}^\dagger(x)  {\cal O} (0) |0\rangle}=   \frac{ \langle {\cal O}(p) |  \widetilde  O_w |{\cal O}(p)\rangle} { \langle {\cal O}(p )|  {\cal O}(p)\rangle} ,
\end{equation}
where $\widetilde  O_w$ is chosen to ensure infrared safety. In general, $\widetilde  O_w$ is a non-time-ordered product of a set of local operators, which again necessitates the use of Wightman functions \cite{Belitsky:2013xxa,Belitsky:2013bja,Belitsky:2013ofa}.  Again, this can be recast as a forward discontinuity.
}
A similar analysis holds for higher order correlators $\langle \widetilde  O_w(1) \widetilde O_w(2)\cdots \rangle$, leading to 
\begin{equation}
\sigma_w= \sum_{c_1,c_2,\cdots}   \int {d^4p_{c_1}} {d^4p_{c_2}}\cdots \, \frac{1}{2i}\,  w(p_{c_1},p_{c_2},\cdots) {\rm Disc}_{M^2} \, T_{\gamma^*c_1'c_2'\cdots\,\rightarrow \,{\gamma'}^*c_1c_2\cdots}\, . \vspace{-10pt}
\end{equation}
\paragraph{Holographic Inclusive Cross Sections} Within the AdS/CFT correspondence, scattering amplitudes can be computed via a perturbative sum of ``Witten diagrams" in analogy to conventional field theory.  Here a 4D, flat space, boundary conformal field theory amplitude is related to a 10D, curved space, bulk string amplitude.  Scattering amplitudes can be written as a bulk amplitude, connected to a boundary CFT function via a convolution, over the curved space, with wavefunctions $\phi(z)$. We find that the differential inclusive cross section for $a+b\rightarrow c +X$ becomes 
\begin{equation} \label{eq:ads6pt}
\frac{d \sigma_{ab\rightarrow c+X}}{d^3{\bf p}_c/E_c} \simeq \frac{1}{2i s} \, \int   \{\Pi_{i=1-6} d\mu (z_i) \phi_n (z_i)\} \,{\rm Disc}_{M^2>0} \{{\cal T}_{abc'\rightarrow a'b'c}(p_i,z_i)\} \, . 
\end{equation}
 
As in conventional QCD, high energy scattering can be dominated by the exchange of Reggeons. Bulk amplitudes, in a AdS space of curvature R, depend on red-shifted external momenta $p^{\mu}$: $\tilde{p}^{\mu}\simeq (z/R)p^{\mu}$.  The dominant contribution, the BPST Pomeron~\cite{Brower:2006ea}, can be described via a propagator with a discontinuity in  $\widetilde s$, with its leading behavior given by $\Disc_{s} \widetilde{\cal K}_P\left(\widetilde s, 0,z,z'\right) \propto {\widetilde s}^{j_0}\,$ , 
with $j_0\simeq 2-2/\sqrt \lambda$\footnote{This is the leading AdS Pomeron intercept.  Higher order corrections can be found in~\cite{Brower:2014wha}}. In the  particular case of DIS, a moment expansion leads to an anomalous dimensions for $j\simeq 2$ at strong coupling given by $
\gamma(j) = \sqrt{2\sqrt \lambda( j-j_0)} -j  + O(\lambda^{-3/4})$.

In the high energy limit, the appropriate 6-point amplitude is dominated by double Pomeron exchange. The amplitude can be written in a factorized form, $T_{abc' \rightarrow a'b'c}=\Phi_{13}*\widetilde{\cal K}_P*V_{c\bar{c}}*\widetilde{\cal K}_P*\Phi_{24}$.  With this factorized form, the kernel and initial wave-function dependence can be integrated out leading to an inclusive particle density $\rho$ for central production given by 
\begin{equation}
\rho(\vec p_T, y,s) \equiv \frac{1}{\sigma_{total}}\frac{d^3\sigma_{ab\to c+X}}{d\mathbf{p}_c^3/E} = \beta \, \int_0^\zm \frac{dz_3}{z_3} \,  \tilde \kappa^ {j_0}\,   [\phi_c(z_3)]^2 \left[{\rm Im} \,  {\cal V}_{c\bar c}\left(\t{\k},0,0\right)\right], \label{eq:disc3t6}
\end{equation}
where  $\beta$ is an overall constant and $\kappa=(-t)(-u)/M^2$. $z_s$ can be determined via the string constant by demanding $2\alpha' \widetilde \k = O(1)$, so that $z_s \sim \frac{R}{\sqrt{2\alpha'\k}}=  \frac{\lambda^{1/4}}{\sqrt{2\k}}$.

We can thus approximate Eq. (\ref{eq:disc3t6}) by integrating only up to $z_3 = z_s<< z_{max}$, where the exponential factor is of order one and can be neglected. In the appropriate kinematic regime, $\phi(z)\simeq z^{\tau}$, where $\tau$ is the twist.  Thus Eq. (\ref{eq:disc3t6}) becomes 
\begin{equation}
\frac{1}{\sigma_{total}}\frac{d^3\sigma_{ab\to c+X}}{d\mathbf{p}_c^3/E_c} = \beta \int_0^{z_{s}} \frac{dz}{z} z^{2\tau_c}   ( \kappa z^2/R^2)^{j_0} e^{-(2  \kappa/\lambda^{1/2})  z^2} \simeq {\beta' }\, \kappa^{-\tau_c}, 
\end{equation}
where we have introduced a new normalization constant $\beta'$. This is the form of Eq.(\ref{eq:Optical6}). For scalar glueballs,  $\tau_c=\Delta_c  = 4$ and we have $\rho(p_\perp, y, s)\sim p_\perp^{-8}$. This result follows essentially from conformality and does not depend on the details of a confinement deformation. It serves as a generalized scaling law  for inclusive distribution, as was worked out for exclusive fixed-angle scattering~\cite{Brodsky:1973kr,*Brodsky:1974vy,*Matveev:1973ra}.  

\section{Conformal Central Prodcution at the LHC}\label{sec:conf}
\begin{figure*}[ht]
\vspace{-10pt}
\begin{center}
 \subfigure[]{
     \includegraphics[scale=.3]{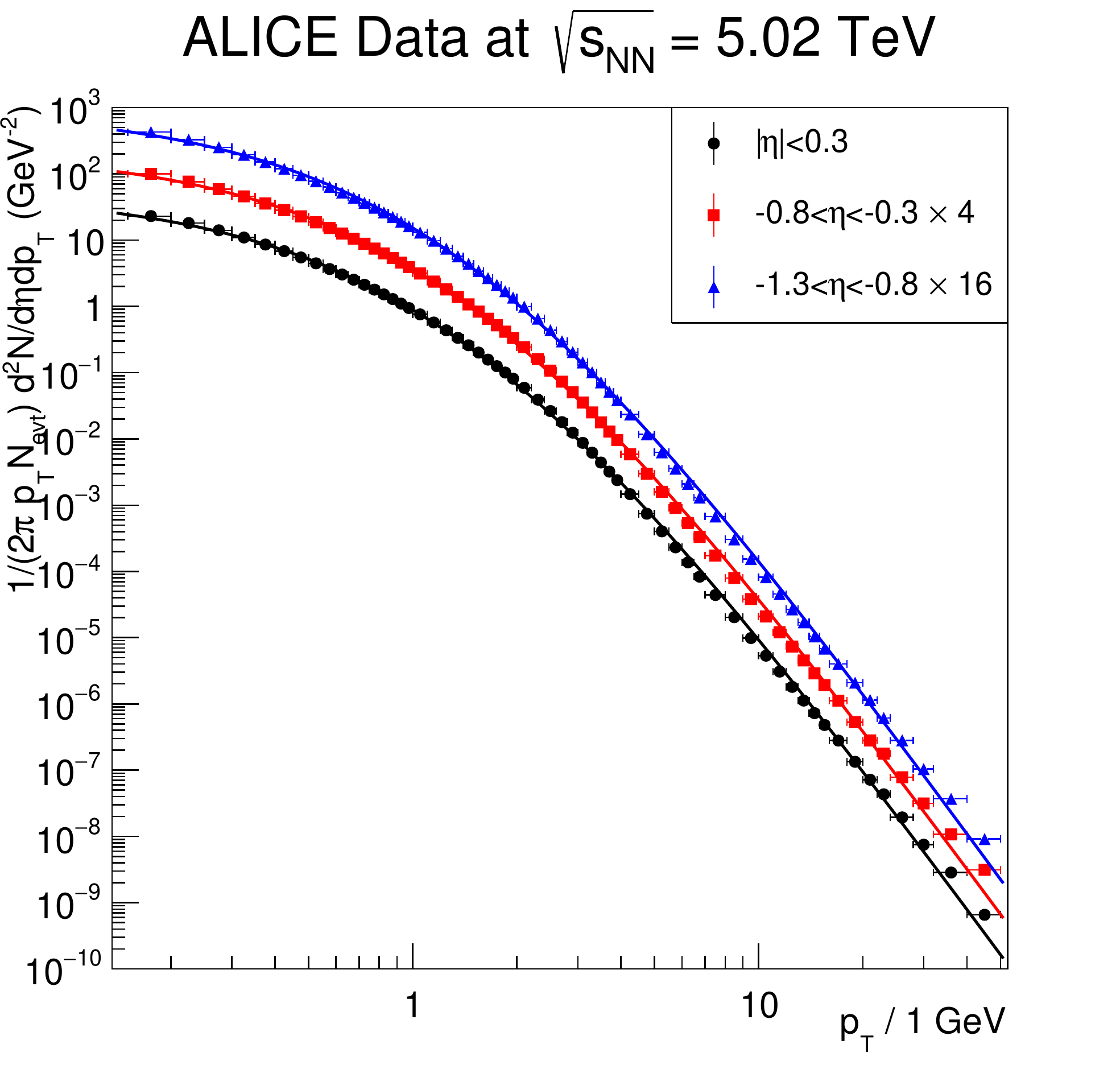}
 \label{fig:1405plot}  
  }\hspace{15pt}
 \subfigure[]{
     \includegraphics[scale=.3]{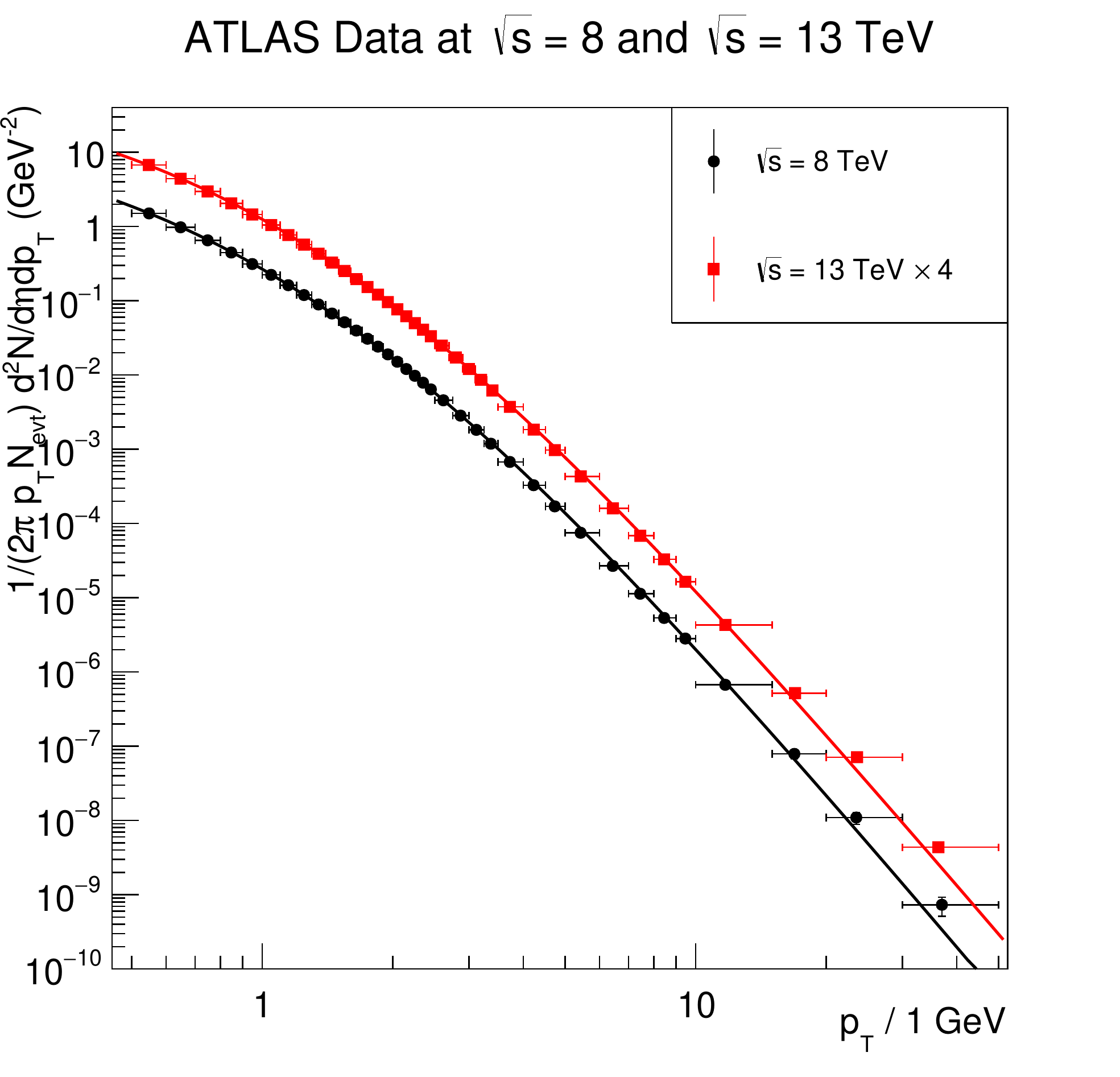}
  \label{fig:atlasfigs} 
  }   
\end{center}
\vspace{-20pt}
\caption[]{\small Fits of inclusive double-differential cross sections. Datasest have been rescaled for visual clarity. \normalsize} 
\label{fig:figs}
\end{figure*}

Phenomenologically we can look for the behavior of Eq.(\ref{eq:Optical6}) via an ansatz: \\$(1/2\pi p_T)(d^2\sigma/d\pt d\eta) = A/(p_T+C)^B$. We consider three data sets: ALICE collab. p-pb collisions at $\sqrt{s_{NN}} $ = 5.02 TeV~\cite{1405}, and ATLAS Collab. p-p collisions at $\sqrt{s}$ = 8 \& 13 TeV~\cite{Aad:2016xww,Aad:2016mok} TeV. The results of fits to our model are shown in Table \ref{tab:fits} and Figures \ref{fig:1405plot} and \ref{fig:atlasfigs}. The ALICE datasets have been run at various pseudorapidity, $\eta$, ranges and demonstrates virtually no variation in kinematics under changes in pseudorapidity.  Overall there is excellent agreement between the fit model and the data. 

\begin{table}[ht]
\begin{centering}
\begin{tabular}{|c|c|c|c|}\hline
Dataset & A/10 (GeV$^{-2}$) & B  & C/(1 GeV)  \\\hline
ALICE  $\left|\eta\right|<0.3$ \cite{1405} & 38.48 $\pm$ 8.26 & 7.23 $\pm$ 0.09 & 1.32 $\pm$ 0.04\\\hline
ALICE  $-0.8 < \eta < -0.3$  \cite{1405} & 37.60 $\pm$ 7.97 & 7.22 $\pm$ 0.08 & 1.30 $\pm$ 0.04  \\\hline
ALICE  $-1.3 < \eta < -0.8$  \cite{1405} & 43.00 $\pm$ 9.29 & 7.30 $\pm$ 0.09 & 1.31 $\pm$ 0.04  \\\hline
ATLAS 8 TeV \cite{Aad:2016xww} & 4.46 $\pm$ 2.60 & 7.03 $\pm$ 0.264 & 1.07 $\pm$ 0.123 \\\hline
ATLAS 13 TeV \cite{Aad:2016mok} & 5.77 $\pm$ 3.38 & 6.96 $\pm$ 0.265 & 1.12 $\pm$ 0.126  \\\hline
\end{tabular}
\caption{Fitted values of $A/(p_T +C)^B$ for three data sets. Both central values and statistical errors are quoted.}
\label{tab:fits}
\end{centering}
\end{table}

The fits are compatible at the two-$\sigma$ level with the power law exponent being independent of both the pseudorapidity and the center of mass measurement. This agrees well with Eq.\ref{eq:Optical6} applied to the AdS/CFT. There are two important caveats, however. First, the overall normalization of the distributions varies sharply between the two types of measurements, with the proton-lead collisions seeming to have a cross section enhanced by an order of magnitude relative to the proton-proton collisons. The holographic argument presented here does not offer an easy way to compute this prefactor, so we have no real prediction for it. Certainly we expect higher-order corrections, which are unaccounted for in our tree-level calculation, to importantly influence the normalization. Moreover, from considerations of the mechanisms for proton-lead and proton-proton scattering, it is clear that the difference between these two can have a physical interpretation, rather than being interpreted as an artifact of our calculation. It is of note that the exact scaling behavior deviates from the expected $\delta=8$ prediction. However this can be explained via the addition of tensor glueball exchange, finite coupling effects, or eikonalization as discussed in~\cite{Nally:2017nsp}.

\FloatBarrier

\begingroup
\bibliographystyle{unsrt}
    \linespread{1}\selectfont
    \bibliography{Inclusive}
\endgroup
\end{document}